\documentclass[11pt]{article}
\usepackage{moriond,epsfig}
\usepackage{wrapfig}

\def\Journal#1#2#3#4{{#1} {\bf #2} (#3) #4}

\def\NP{\em Nucl. Phys.}

\def\PL{\em Phys. Lett.}

\def\ZP{\em Z. Phys.}

\def\JP{\em J. Phys.}
\def\EPJ{\em Eur. Phys. J.}
\def\CPC{\em Comp. Phys. Comm.}

\def\JHEP{\em JHEP}
\def\etal{{\it et al.}}
\def\ibid{{\it ibid.}}

\def\be{\begin{equation}}
\def\ee{\end{equation}}
\def\bea{\begin{eqnarray}}
\def\eea{\end{eqnarray}}

\newcommand{\rs}		{\ensuremath{\sqrt{s}}}

\newcommand{\oaa}		{\ensuremath{\mathcal{O}(\alpha_\mathrm{S}^2)}}
\newcommand{\oaaa}		{\ensuremath{\mathcal{O}(\alpha_\mathrm{S}^3)}}
\newcommand{\as}		{\ensuremath{\alpha_\mathrm{S}}}
\newcommand{\asrs}		{\ensuremath{\alpha_\mathrm{S}(\sqrt{s})}}
\newcommand{\asmz}		{\ensuremath{\alpha_\mathrm{S}(M_{\mathrm{Z^0}})}}
\newcommand{\chisq}	{\ensuremath{\chi^2}}
\newcommand{\chisqd}	{\ensuremath{\chi^2/\mathrm{d.o.f.}}}
\newcommand{\xmu}		{\ensuremath{x_{\mu}}}
\newcommand{\cp}		{\ensuremath{C}}
\newcommand{\bt}		{\ensuremath{B_T}}
\newcommand{\bw}		{\ensuremath{B_W}}
\newcommand{\mh}		{\ensuremath{M_H}}

\newcommand{\thr}		{\ensuremath{1-T}}
\newcommand{\yy}		{\ensuremath{y_{23}}}
\newcommand{\gev}		{\ensuremath{\mathrm{GeV}}}
\newcommand{\py}		{{\sc Pythia}}
\newcommand{\hw}		{{\sc Herwig}}
\newcommand{\ar}		{{\sc Ariadne}}
\newcommand{\jt}		{{\sc Jetset}}
\newcommand{\cj}		{{\sc Cojets}}
\newcommand{\pyy}		{{\sc Pythia~5.7}}
\newcommand{\hww}		{{\sc Herwig~5.9}}
\newcommand{\arr}		{{\sc Ariadne~4.08}}

\newcommand{\jttold}	{{\sc Jetset~6.3}}
\newcommand{\epem}		{\ensuremath{\mathrm{e^+e^-}}}

\newcommand{\znull}	{\ensuremath{\mathrm{Z^0}}}
\newcommand{\lnr}		{\ensuremath{\ln(R)}}
\newcommand{\T} 		{\displaystyle}

\begin{document}
\vspace*{4cm}
\title{DETERMINATIONS OF \as\ AT \rs~= 14 TO 44~GEV \\ USING RESUMMED CALCULATIONS}

\author{ P.A. MOVILLA FERN\'{A}NDEZ}

\address{Max-Planck-Institut f\"ur Physik, F\"ohringer Ring 6, 
D-80805 M\"unchen, Germany \\ (pedro@mppmu.mpg.de)}

\maketitle\abstracts{ 
The strong coupling constant \as\ is determined using recently
re-analysed \epem\ annihilation data collected by the JADE experiment
at \rs~=14 to 44~GeV. The measurements are based on \oaa+NLLA
predictions for various event shape observables. The calculations are
found to describe reliably data at the lowest energies of the \epem\
continuum where non-perturbative contributions become important. The
results for \as\ are in good agreement with the QCD expectation for
the running of the strong coupling constant.  This is the first
determination of \as\ at \rs~= 14 and 22~GeV based on resummed QCD
predictions.}

\section{Introduction}\label{sec:intro}
Tests of Quantum Chromodynamics (QCD) substantially benefit from
\epem\ annihilation experiments at lower centre-of-mass energies \rs\
since the characteristic energy evolution of the theory is expected to
become more manifest towards decreasing \rs. The re-analysis of data
collected by the JADE experiment~\cite{JADE-alphas,JADE-etc}, as
counterpart to the LEP data, has been shown to be a valuable effort.
Recently, data at energies down to \rs~= 14~GeV could be employed in
state-of-the-art QCD studies due to the successful resurrection of the
original JADE simulation and event reconstruction software.

In the last decade, significant progress has been made in the
theoretical calculations of event shape observables serving as
powerful tools to investigate perturbative and non-perturbative
aspects of QCD particularly at PETRA energies.  This analysis focuses
on \as\ determinations based on resummed calculations for event shapes
which are applied here for the first time at \rs~= 14 and
22~GeV. Event shape data have also been used to assess the performance
of various QCD event generators tuned to LEP data.

\section{Data Samples and MC Simulation}\label{sec:data}

\begin{wrapfigure}[32]{r}{85mm}
\vspace*{-16mm}
\hspace*{-3mm}\psfig{file=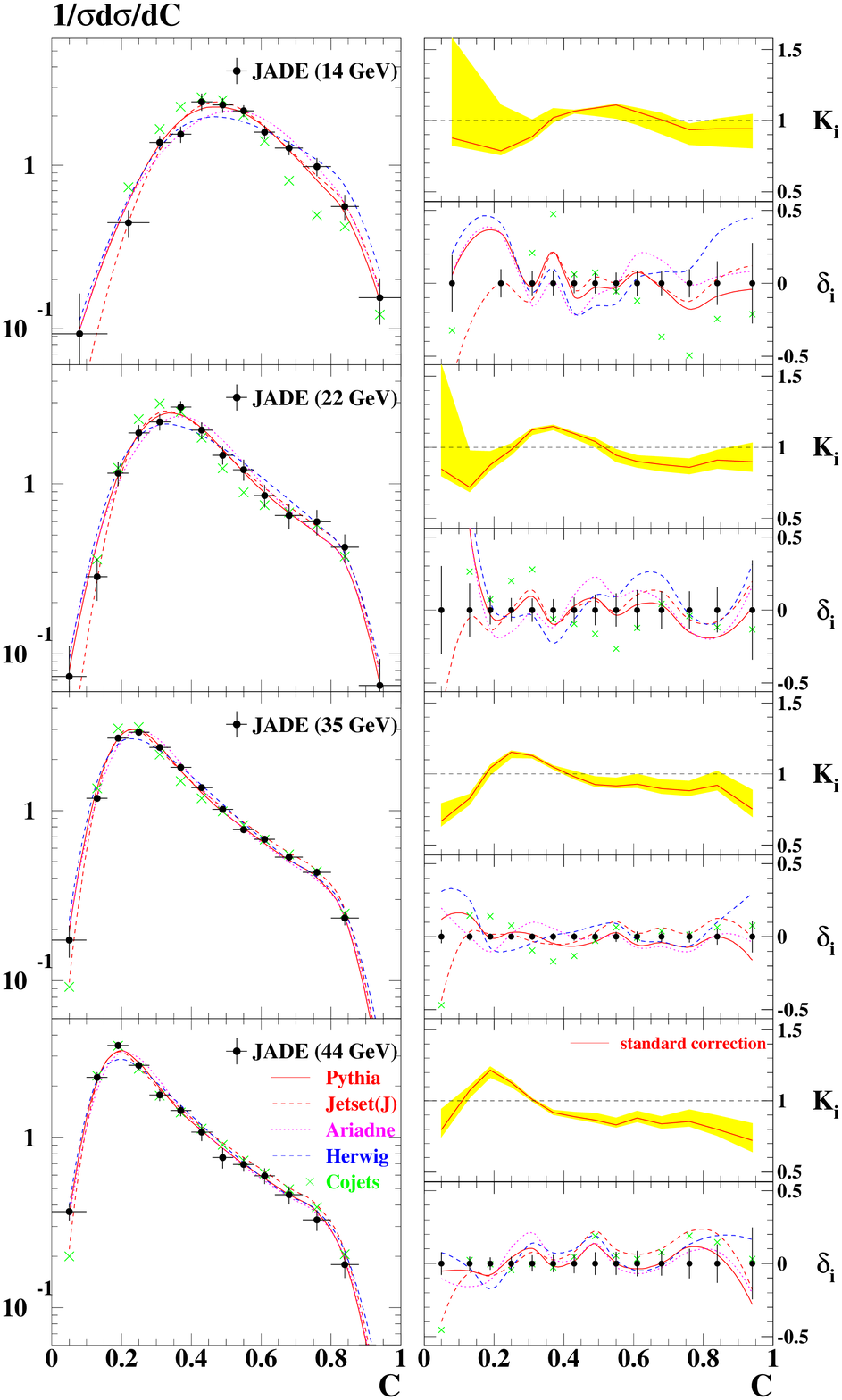,width=90mm,clip=}
\vspace*{-9mm}
\caption {\label{fig:had} 
Hadron level distributions for \cp\ at \rs~= 14 to 44~GeV compared
with the predictions of various QCD event generators (left).  The
error bars denote the total errors. Also shown are the detector
correction factors $K_i$ and the normalised differences $\delta_i$
between model and data (right).}
\end{wrapfigure}

The JADE detector~\cite{JADE-Exp} was operated from 1979 to 1986 at
the PETRA \epem\ collider. It was designed as a hybrid $4\pi$-detector
to measure charged and neutral particles.  The studies presented here
are based on multihadronic data samples (with number of events in
brackets) at \rs~= 14.0 (1734), 22.0 (1390), 34.6 (14372, data taken
1981-82), 35.0 (20688, data taken 1986), 38.3 (1587) and 43.8~GeV
(3940). Simulated data were generated using the QCD event generators
\pyy, \arr\ and \hww~\cite{QCDMC} combined with the JADE detector
simulation. We adopted the parameter sets used by the OPAL
experiment~\cite{OPALtune} to describe \epem\ data at \rs~=
$M_\znull$.  We also considered a predecessor version
\jttold~\cite{QCDMCold} used in former JADE studies since it was shown
to describe \epem\ hadronic final states. Comparisons of the simulated
and measured distributions of various integral and spectral quantities
generally gave a good description of the measured data.  So the
simulation can be used to correct for detector effects.

\section{Event Shapes at PETRA Energies}\label{sec:shapes}

From the data passing the multihadronic selection
criteria~\cite{JADE-Exp}, the distributions of the event shape
observables {\em thrust} \thr, {\em heavy jet mass} \mh, {\em total}
and {\em wide jet broadening} \bt\ and \bw, {\em \cp\ parameter} and
the {\em differential 2-jet rate}
\yy\ based on the Durham scheme are calculated
(cf.~\cite{JADE-alphas}). The distributions are corrected for the
limited acceptance and resolution of the detector and for initial
state photon radiation effects using \py\ for the standard correction.
Since mass effects due to the electroweak decay of heavy b-hadrons
faking gluon activity in the 3-jet region are crucial at \rs~= 14 and
22~GeV, we take the contribution $\epem\to\mathrm{b\bar{b}}$ as an
additional background to be subtracted from the distributions.

As an example, the resulting {\em hadron level} data distributions for
\cp\ are represented by Fig. \ref{fig:had}. For comparison, the
respective distributions predicted by the QCD models based on u-, d-,
s- and c-flavoured events are shown. In case of \py, there is
generally a good agreement between the data and the model over the
whole kinematic range of the observables. The performance of \ar\ and
\hw\ is more moderate at 14~GeV and improves at increasing
c.m.s. energies. \hw\ significantly underestimates the peak region of
the distributions. In contrast, the JADE-based \jt\ version fits the
14~GeV data but increasingly deviates from the data at higher
energies. The prediction of \cj\ \cite{QCDMCcoj} is clearly
disfavoured by the lower energy data and remains worse also at higher
energies. Thus we do not consider this model for our \as\ studies.

\begin{figure}[t]
\vspace*{-2mm}
\hspace*{-8mm}
\parbox{85mm}{
\psfig{file=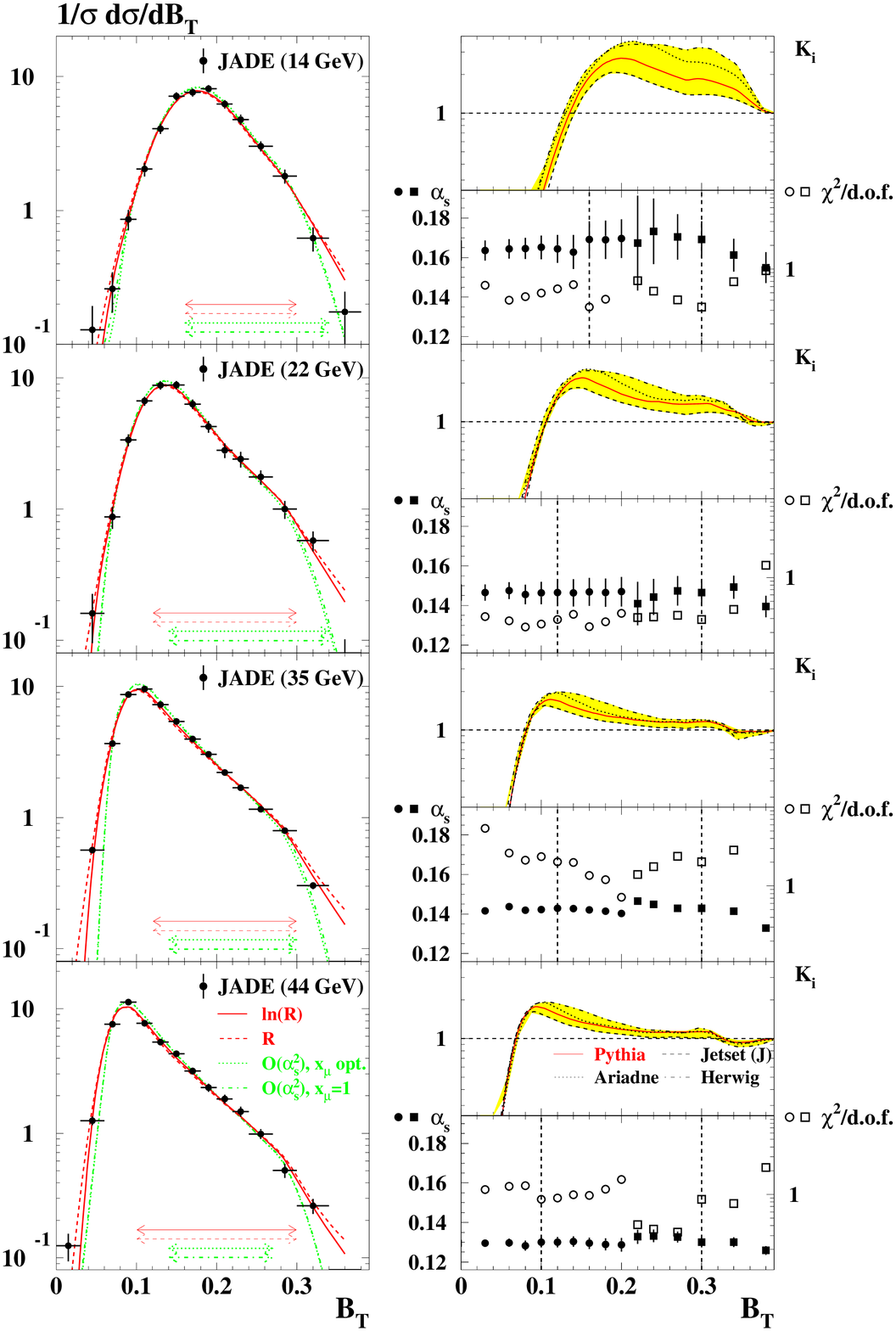,width=85mm,clip=} \vspace*{-9mm}
\caption{\label{fig:fits} 
Fits of \oaa+NLLA to \bt\ at \rs~= 14 to 44~GeV (left). Hadronisation
corrections $K_i$ and model uncertainties are shown as well as the
dependence of the results for \as\ and the
\chisqd\ on the variation of the fit range (right).}}
\hspace*{1mm}
\parbox[c][142mm][b]{79mm}{\hspace*{-4mm}
\psfig{file=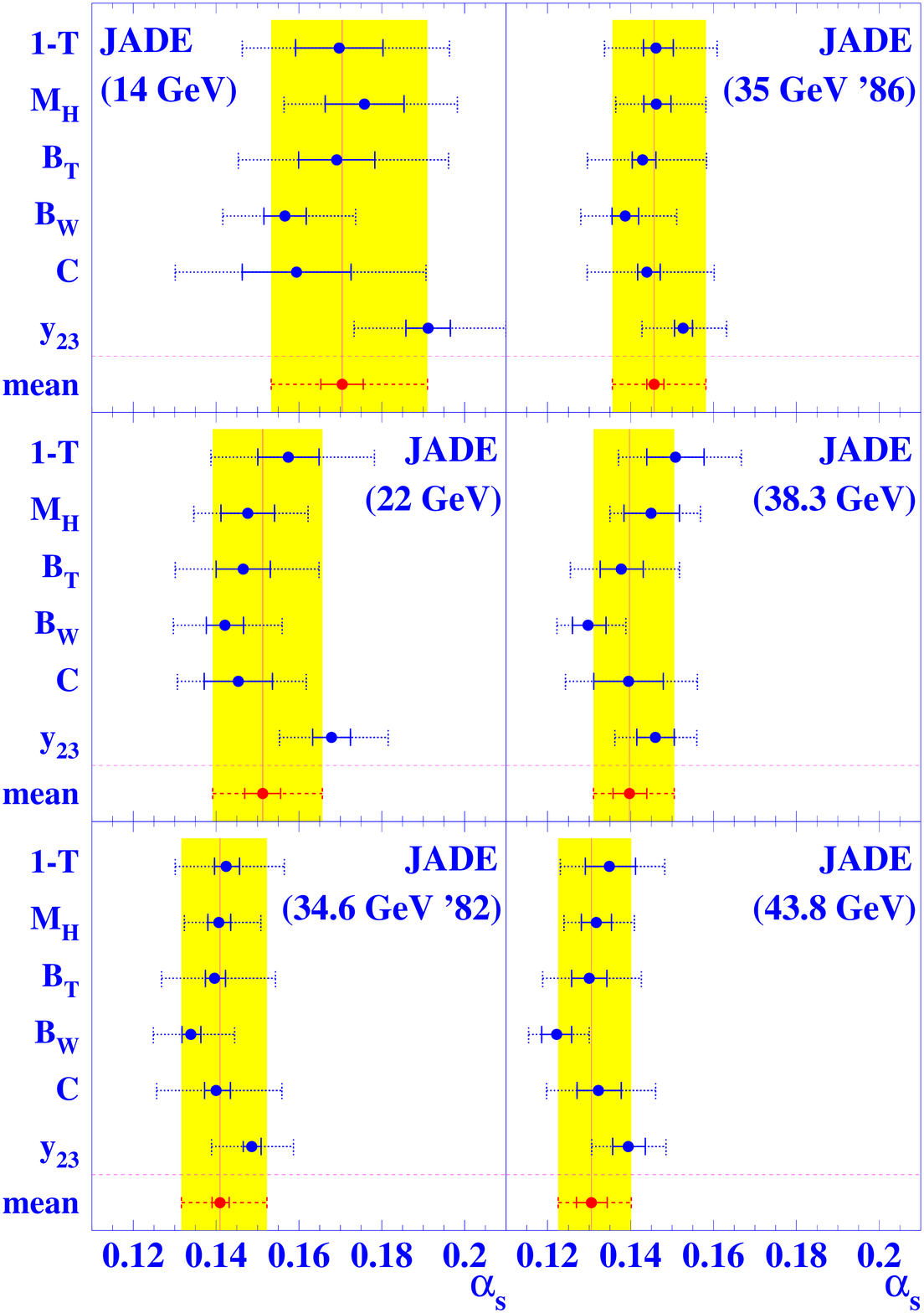,width=85mm,clip=} \vspace*{-9mm}
\caption{\label{fig:asresults} 
Results for \asrs\ at \rs~= 14 to 44 GeV derived from \oaa+NLLA fits to
\thr, \mh, \bt, \bw, \cp\ and \yy.  The inner error bars denote
experimental uncertainties, the outer error bars are the total
errors.}}
\vspace*{-3mm}
\end{figure}

\section{Measurements of \as}\label{sec:alphas}

The determination of \as\ is based on a combination of an exact QCD
matrix element calculation \oaa~\cite{ERT} intended to describe the
3-jet region of phase space and a next-to-leading-logarithmic
approximation~\cite{NLLA} (NLLA) valid in the 2-jet region where
multiple radiation of soft and collinear gluons from a system of two
hard back-to-back partons dominate.  We perform
\chisq-fits of the theoretical predictions corrected for
 hadronisation effects.  For the main results, we use the
\lnr-matching~\cite{NLLA} for the perturbative prediction with 
the renormalisation scale factor \xmu~$\equiv\mu/\rs$~=~1 and \py\ for
the estimation of non-perturbative contributions.  As an example,
Fig.~\ref{fig:fits} shows the fitted predictions for \bt. We generally
observe stable fits and good agreement with the data at all
c.m.s. energies with \chisqd\ ranging from about 0.2 to 2.0. In case
of \bw, a significant excess of the theory over the data in the 3-jet
region of the distributions is present.

In principle we follow the procedure in~\cite{JADE-alphas} to estimate
experimental and theoretical systematic errors but include additional
MC modelling uncertainties.  Hadronisation effects and uncertainties
increase significantly at 14~GeV. Experimental errors are under
control for all data samples.  On the basis of fit and experimental
errors, the individual results are consistent with each other within
1-2 standard deviations. For each c.m.s. energy, the
\as-results of the six observables were combined using the weighted
mean method~\cite{JADE-alphas}.  The final results obtained are listed
in Tab.~\ref{tab:asresults}.  The total errors are dominated by higher
order uncertainties. At \rs~= 14 and 22~GeV, hadronisation uncertainties are
very large but still of the same order as the renormalisation scale
uncertainties.

\section{Conclusions}

\begin{wrapfigure}[15]{r}{80mm}\vspace*{-17mm}\hspace*{-2.5mm}
\psfig{file=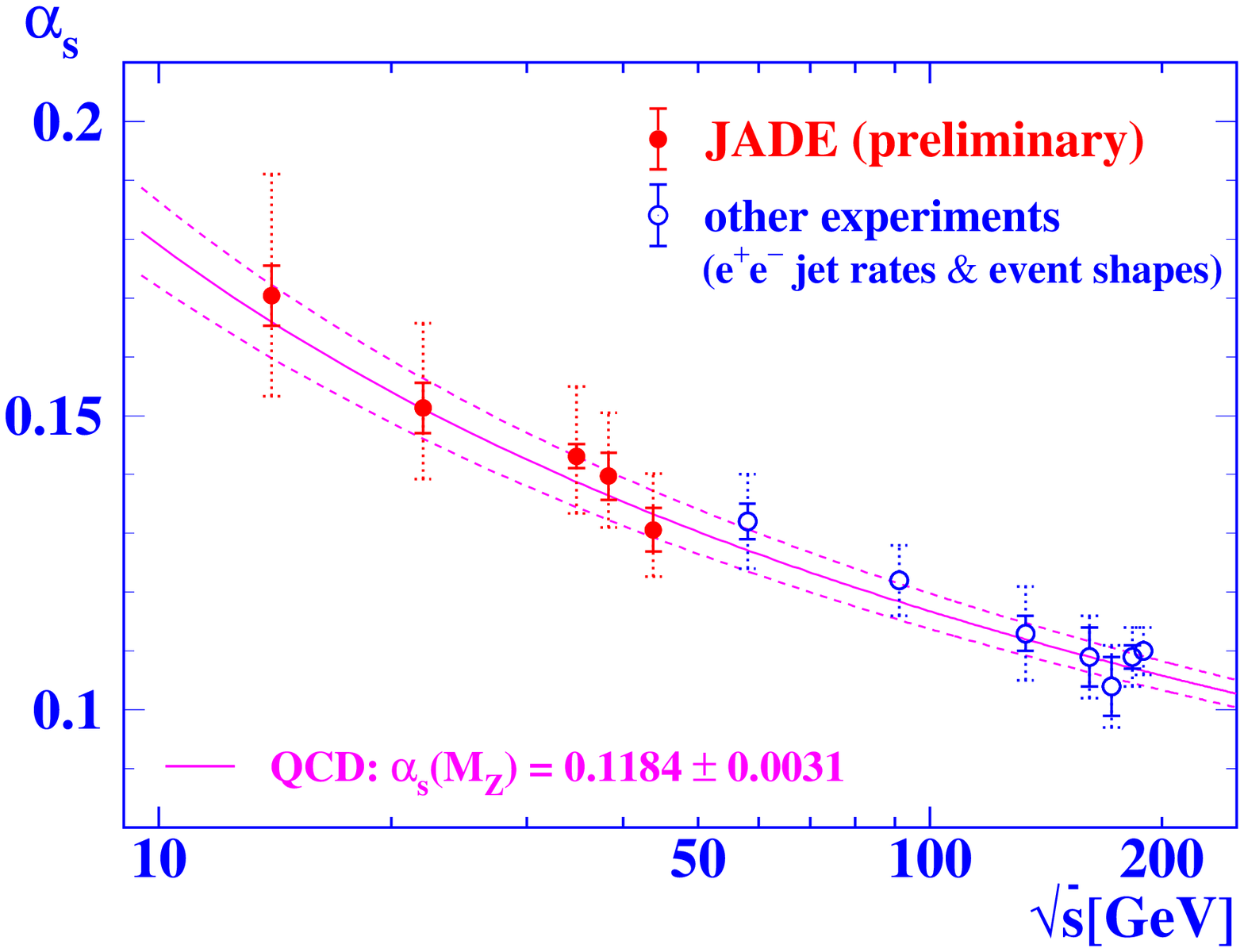,width=90mm,clip=}
\vspace*{-10mm}
\caption{\label{fig:asrunning} 
\as-values derived from this analysis compared with 
corresponding results at higher energies. Also shown is the QCD
expectation for the world average of \as~\protect\cite{Bet00}.}
\end{wrapfigure}
Resummed QCD theory fits event shape data measured with JADE well down
to \rs~= 14~GeV and allow consistent determinations of \as. The values
obtained at the lowest energies are affected by large hadronisation
uncertainties. The LEP-tuned \py\ Monte Carlo used for the estimation
of non-perturbative effects is surprisingly well capable of describing
many aspects of \epem\ hadronic final states at PETRA energies. The
\as\ results obtained here as well as in similar analyses at higher energies
based on resummed event shapes (Fig. \ref{fig:asrunning}) agree well
with the QCD expectation for the running coupling~\cite{Bet00}. A
\chisq\ fit of the \oaaa\ prediction to these values taking only
experimental errors into account yields \asmz= $0.1213 \pm 0.0006$
with \chisqd=$8.3/11$. Even considering total errors, the hypothesis
of a constant value of \as\ is disfavoured by a fit probability of
$\approx 10^{-5}$. The JADE data significantly improve the
verification of QCD on basis on \epem\ data. \par

\vspace*{-1mm}
\begin{table}[h] \small
\caption{\label{tab:asresults} 
 Preliminary results for \as\ derived from the individual results
 using the weighted average method.}
\vspace*{2mm}
\begin{center}
\renewcommand{\arraystretch}{1.4}
\begin{tabular}{|l|c|cc|c|c|c|}  \hline
\rs\ $[\gev]$ & 
  \as(\rs) & fit
	     & exp.
		& hadr.
		   &  higher ord.
			& total \\ \hline\hline
     14.0      &   $0.1704$ &   \multicolumn{2}{c|}{$\pm 0.0051$} 
	& $\T +0.0141 \atop\T -0.0136$ & $\T +0.0143 \atop\T -0.0091$ & $\T +0.0206\atop\T -0.0171$ \\  
     22.0      &   $0.1513$ &   \multicolumn{2}{c|}{$\pm 0.0043$} 
	& $\pm 0.0101                $ & $\T +0.0101 \atop\T -0.0065$ & $\T +0.0144\atop\T -0.0121$ \\  
     34.6 ('82)&   $0.1409$ &   $\pm 0.0012$ & $\pm 0.0017$      
	& $\pm 0.0071                $ & $\T +0.0086 \atop\T -0.0057$ & $\T +0.0114\atop\T -0.0093$ \\  
     35.0 ('86)&   $0.1457$ &   $\pm 0.0011$ & $\pm 0.0020$      
	& $\pm 0.0076                $ & $\T +0.0096 \atop\T -0.0064$ & $\T +0.0125\atop\T -0.0101$ \\  
     38.3      &   $0.1397$ &   $\pm 0.0031$ & $\pm 0.0026$      
	& $\pm 0.0054                $ & $\T +0.0084 \atop\T -0.0056$ & $\T +0.0108\atop\T -0.0087$ \\  
     43.8      &   $0.1306$ &   $\pm 0.0019$ & $\pm 0.0032$      
	& $\pm 0.0056                $ & $\T +0.0068 \atop\T -0.0044$ & $\T +0.0096\atop\T -0.0080$ \\  
\hline
\end{tabular}\end{center}
\end{table}

\end{document}